# Universal gapless Dirac cone and tunable topological states in $(MnBi_2Te_4)_m(Bi_2Te_3)_n$ heterostructures

Yong Hu[1,2,#], Lixuan Xu[3,#], Mengzhu Shi[1,2,#], Aiyun Luo[4,#], Shuting Peng[1,2], Z. Y. Wang[1,2], J. J. Ying[1,2], T. Wu[1,2], Z. K. Liu[3], C. F. Zhang[3,5], Y. L. Chen[3,6], G. Xu[4], X.-H. Chen[1,2,*] and J.-F. He[1,2,*]

[1]Hefei National Laboratory for Physical Sciences at the Microscale and Department of Physics, University of Science and Technology of China, Hefei, Anhui 230026, China

[2]Chinese Academy of Sciences Key Laboratory of Strongly-coupled Quantum Matter Physics, University of Science and Technology of China, Hefei, Anhui 230026, China

[3]School of Physical Science and Technology, ShanghaiTech University and CAS-Shanghai Science Research Center, Shanghai 201210, China

[4]Wuhan National High Magnetic Field Center and School of Physics, Huazhong University of Science and Technology, Wuhan 430074, China

[5]College of Advanced Interdisciplinary Studies, National University of Defense Technology, Changsha 410073, China

[6]Department of Physics, University of Oxford, Oxford, OX1 3PU, UK

#These authors contributed equally to this work.

*Correspondence and requests for materials should be addressed to X.-H.C. (chenxh@ustc.edu.cn) and J.-F.H. (jfhe@ustc.edu.cn).

In the newly discovered magnetic topological insulator $MnBi_2Te_4$ [1-8], both axion insulator state and quantized anomalous Hall effect (QAHE) have been observed by tuning the magnetic structure [9-11]. The related $(MnBi_2Te_4)_m(Bi_2Te_3)_n$ heterostructures with increased tuning knobs, are predicted to be a more versatile platform for exotic topological states [12-16]. Here, we report angle-resolved photoemission spectroscopy (ARPES) studies on a series of the heterostructures ($MnBi_2Te_4$, $MnBi_4Te_7$ and $MnBi_6Te_{10}$). A universal gapless Dirac cone is observed at the $MnBi_2Te_4$ terminated (0001) surfaces in all systems. This is in sharp contrast to the expected gap from the original antiferromagnetic ground state [2-4,9,17-19], indicating an altered magnetic structure near the surface, possibly due to the surface termination. In the meantime, the electron band dispersion of the surface states, presumably dominated by the top surface [20,21], is found to be sensitive to different stackings of the underlying $MnBi_2Te_4$ and $Bi_2Te_3$ layers. Our results suggest the high tunability of both magnetic and electronic structures of the topological surface states in $(MnBi_2Te_4)_m(Bi_2Te_3)_n$ heterostructures, which is essential in realizing various novel topological states [16,22].

Topological materials with intrinsic magnetic orders are believed to be an ideal platform to explore and manipulate exotic topological quantum states [23-31]. This idea is recently realized in a stoichiometric magnetic topological insulator $MnBi_2Te_4$ [3,9-11]. The inherent antiferromagnetic order is predicted to induce gapped surface states, which can host an exotic axion insulator state with topological magnetoelectric effect (TME) (Fig. 1a) [2,3]. Recent transport results further indicate that the axion insulator state can be transformed into Chern insulator state with QAHE by tuning the magnetic structure into ferromagnetism under external field (Fig. 1a) [9-11]. However, unexpected gapless topological surface states (TSSs) are observed by high-resolution ARPES measurements [32-35], raising critical questions regarding the surface magnetic structure, magnetic coupling strength and the specifics of the $MnBi_2Te_4$ material system.

$(MnBi_2Te_4)_m(Bi_2Te_3)_n$ heterostructures represent a closely related but more complex system, in which both $MnBi_2Te_4$ and $Bi_2Te_3$ layers are essential building blocks (Fig. 1b,c) [1,2,14-16]. The reduced interlayer antiferromagnetic exchange coupling (due to the nonmagnetic $Bi_2Te_3$ layers) facilitates the magnetic manipulation [14-16], and two distinct building blocks provide an additional knob to tune the topological states by different stackings of the van der Waals layers (Fig. 1b,c). In this sense, the $(MnBi_2Te_4)_m(Bi_2Te_3)_n$ heterostructures are predicted to be a more versatile playground for realizing exotic topological states and topological phase transitions [14-16].

We have performed ARPES studies on a series of $(MnBi_2Te_4)_m(Bi_2Te_3)_n$ heterostructures, including $MnBi_2Te_4$ (m=1, n=0), $MnBi_4Te_7$ (m=1, n=1) and $MnBi_6Te_{10}$ (m=1, n=2). A gapless Dirac cone is observed on the $MnBi_2Te_4$ terminated (0001) surfaces in all systems, demonstrating a universal gapless surface state in the $(MnBi_2Te_4)_m(Bi_2Te_3)_n$ family. This is consistent with an altered magnetic order on the cleaved $MnBi_2Te_4$ surface, echoing the tunability of the magnetic structure in the system. The band dispersions of the TSSs are also examined in these heterostructures. Different from the surface states in typical van der Waals topological insulators, whose band dispersions are mainly determined by the surface layer itself [20,21], the surface states in $(MnBi_2Te_4)_m(Bi_2Te_3)_n$ are found to be sensitive to different stackings of the underlying $MnBi_2Te_4$ and $Bi_2Te_3$ layers. Such a behavior, when combined with the tunable magnetic structure, provides an experimental basis for the theoretical proposal to realize exotic topological states by manipulating the magnetic structure and different combinations of the two building block layers in $(MnBi_2Te_4)_m(Bi_2Te_3)_n$ heterostructures [16,22].

When the as-grown single crystalline heterostructures are cleaved for ARPES measurements, different terminations are expected due to the similar strength of the van der Waals bonds in the material. By using a small spot laser-based ARPES system (see method for details), we have resolved different terminations in the real space of the cleaved sample surface and probed their electronic structures, respectively. In $MnBi_6Te_{10}$, three types of electronic structures are identified (Fig. 2), evidenced by three sets of constant energy maps (Fig. 2b,e,h) and electron bands (Fig. 2c,f,i and supplementary Fig. S1). This is consistent with the three possible terminations of the $MnBi_6Te_{10}$ system (Fig. 2a,d,g; see supplementary Fig. S2 for details of the correspondence between the electronic structure and termination). The same is true for the $MnBi_4Te_7$ ($MnBi_2Te_4$) system, in which two (one) types of electronic structures are found (see supplementary Fig. S3).

We first focus on the $MnBi_2Te_4$ termination, which is the essential magnetic building block of all $(MnBi_2Te_4)_m(Bi_2Te_3)_n$ heterostructures. Comparative measurements have been carried out on the $MnBi_2Te_4$ termination of $MnBi_2Te_4$, $MnBi_4Te_7$ and $MnBi_6Te_{10}$ systems below their magnetic transition temperatures, respectively (Fig. 3). A Dirac-like TSS is observed around the $\Gamma$ point of all three materials (marked by the orange dashed lines in Fig. 3d,h,l). Two branches of the dispersion intersect each other at a single Dirac point, showing the gapless nature of the TSS. This is in sharp contrast to the earlier reports of a magnetic order induced surface gap in the order of hundred-meVs in both $MnBi_2Te_4$ [2,4,9,17] and $MnBi_4Te_7$ [14,15]. On the contrary, our results are consistent with the recent observation of a gapless TSS in $MnBi_2Te_4$ [32-35]. We note that the gapless TSSs in our experiments persist to high temperatures above the magnetic transition temperatures of the materials (supplementary Fig. S4). The universal Dirac point unveiled in the current study establishes the gapless TSSs as an intrinsic property of the $MnBi_2Te_4$ terminated $(MnBi_2Te_4)_m(Bi_2Te_3)_n$ heterostructures.

After showing the gapless Dirac point, we move to the specific band dispersion of the TSSs (see supplementary Fig. S5 for the identification of the TSSs). Different from the Dirac point whose existence is guaranteed by the topological character of the bulk states, the specific band dispersion of the TSSs is always tied to the surface layer itself [20,21]. For example, van der Waals heterostructures with one quintuple layer (QL) $Bi_2Te_3$ on top of ten QL $Bi_2Se_3$ (or $Sb_2Te_3$) give rise to almost the same surface band dispersion as that in 3D topological insulator $Bi_2Te_3$ [21,36]. Surprisingly, different band dispersions of the TSSs are found on the same $MnBi_2Te_4$ terminated surface layer in our systems (Fig. 3). First, the band of the TSSs away from the Dirac point disperses differently between $MnBi_2Te_4$, $MnBi_4Te_7$ and $MnBi_6Te_{10}$ (e.g. different slopes of the band dispersion, marked by the orange dashed lines in Fig. 3d,h,l). Second, the location of the Dirac point moves to a deeper binding energy in $MnBi_6Te_{10}$ comparing to that in $MnBi_2Te_4$ and $MnBi_4Te_7$ (Fig. 3n). The same is true for the $Bi_2Te_3$ termination, on which different band dispersions of the TSSs are identified (Fig. 4) (see supplementary Fig. S6 for the identification of the TSSs). In $MnBi_6Te_{10}$, two distinct topological surface bands are observed on two types of $Bi_2Te_3$ surface layers, respectively (Fig. 4a-h). These two $Bi_2Te_3$ surface layers are identical by themselves but inequivalent by considering different stackings of the layers beneath them (compare Figs. 4b and 4f). $Bi_2Te_3$ terminated surface also exists in $MnBi_4Te_7$ which shows a topological surface band different from the above two in $MnBi_6Te_{10}$ (Fig. 4 i-l). These results show that the TSSs in $(MnBi_2Te_4)_m(Bi_2Te_3)_n$ heterostructures are sensitive not only to the surface layer, but also to the underlying layers.

Now we discuss the possible origin of the above key observations in $(MnBi_2Te_4)_m(Bi_2Te_3)_n$ heterostructures. The first thing is to understand the universal gapless TSSs. The magnetic moments in the ground state of $(MnBi_2Te_4)_m(Bi_2Te_3)_n$ are ferromagnetically ordered within a Mn-plane along the z-direction, but antiferromagnetically coupled between layers [6,37,38]. In theory, this so-called A-type antiferromagnetic order would inevitably induce a sizable magnetic gap in the (0001) surface states, which is incompatible with the gapless TSSs observed in our experiment. The universality of the gapless TSSs rules out any specific property of the $MnBi_2Te_4$ crystal as the origin. Different strength of the interlayer antiferromagnetic exchange coupling between $MnBi_2Te_4$, $MnBi_4Te_7$ and $MnBi_6Te_{10}$ also makes the interlayer coupling strength less relevant in this connection. One simple explanation involves impurity states, which would potentially fill up the intrinsic magnetic gap. This scenario seems to be consistent with the earlier experimental reports of defects and vacancies [2,4-6]. However, electron density of states from localized impurities cannot give rise to any band dispersion in the momentum space. This makes the impurity scenario hard to explain our experimental observation of a single Dirac point at the intersection of two bands [39]. Another possibility is the existence of a trivial surface state in the magnetic gap. However, we have observed a universal Dirac point with different band dispersions in $MnBi_2Te_4$, $MnBi_4Te_7$ and $MnBi_6Te_{10}$, indicating a topological nature of the Dirac-like surface states. The third possibility is the co-existence of multiple short-range magnetic orders (presumably in different domains), such that the global magnetic moment in each layer could be zero. Transport measurement on $MnBi_4Te_7$ does reveal a competition between different magnetic orders at low temperature – due to the moderate interlayer antiferromagnetic exchange coupling [15]. This scenario can also be naturally extended to $MnBi_6Te_{10}$, which has an even weaker interlayer coupling [15]. However, it contradicts the neutron diffraction measurements on $MnBi_2Te_4$, in which the A-type long-range antiferromagnetic order has been established as the ground state [6]. Therefore, if we assume the same origin for the observed gapless TSSs in $(MnBi_2Te_4)_m(Bi_2Te_3)_n$ heterostructures, a more plausible explanation is the change of magnetic configuration in the few top layers of the sample – possibly during the cleaving process. In $MnBi_2Te_4$, several magnetic structures have been suggested to produce the gapless TSSs (Fig. 3m), including A-type antiferromagnetic order with in-plane magnetic moment, G-type antiferromagnetic order (magnetic moments aligned antiferromagnetically both in plane and between adjacent layers) and disordered magnetic moments (or paramagnetic state) [32]. We have performed density functional theory (DFT) calculations and confirmed that the gapless TSSs can also be induced by these magnetic structures in $MnBi_4Te_7$ and $MnBi_6Te_{10}$. While more efforts are needed to pin down the exact magnetic structure near the surface, we find that the calculated bands from the G-type antiferromagnetic order and paramagnetic state exhibit a better agreement to the experimental results comparing to that from the A-type in-plane antiferromagnetic order (see supplementary Fig. S7). Regardless of the exact new state, our results indicate that the magnetic structure of the $(MnBi_2Te_4)_m(Bi_2Te_3)_n$ system may be easily changed, supporting the proposal to realize various topological quantum phases via magnetic tuning [16,22].

Next, we discuss the band dispersions of the TSSs in $(MnBi_2Te_4)_m(Bi_2Te_3)_n$ heterostructures. In topological materials, the gapless nature of the TSSs is determined by the topological character of the bulk band, but the specific band structure of the TSSs is dominated by the environment near the sample surface [20,21]. This environment is typically determined by the surface layer itself. However, in our measurements, different topological surface band dispersions are observed on nominally the same surface termination, indicating that the surface environment of the $(MnBi_2Te_4)_m(Bi_2Te_3)_n$ is changed by different stackings of the underlying layers − especially the magnetic layers of $MnBi_2Te_4$. This point is also supported by the experimental results. First, the band dispersion of the TSSs on the $Bi_2Te_3$ termination becomes more similar to that of 3D $Bi_2Te_3$ when the $MnBi_2Te_4$ layer is separated away from the surface by another layer of $Bi_2Te_3$ (Fig. 4a-d). This band dispersion is substantially modified when the $MnBi_2Te_4$ layer locates adjacent to the surface (Fig. 4e-h). Second, the band dispersion of the TSSs on the $MnBi_2Te_4$ termination is also significantly changed when the coupling between two $MnBi_2Te_4$ layers is reduced by the intercalated $Bi_2Te_3$ layer(s) (Fig. 3). In this sense, the existence of two building blocks in $(MnBi_2Te_4)_m(Bi_2Te_3)_n$ heterostructures provides a unique tuning knob to manipulate the TSSs without changing the surface of the sample. The direct response between the TSSs and the interlayer coupling also echoes the theoretical proposal to realize various exotic topological phases by different stackings and thicknesses of the $MnBi_2Te_4$ and $Bi_2Te_3$ layers in this system [16]. It would be instructive to quantitatively investigate how the magnetic layers and interlayer coupling modify the surface environment, as well as whether they are related to the possible change of magnetic configuration in the few top layers of the sample.

In conclusion, we have systematically studied a series of $(MnBi_2Te_4)_m(Bi_2Te_3)_n$ heterostructures with different terminations. A universal gapless TSS is established at the $MnBi_2Te_4$ terminated surfaces, being consistent with a changed magnetic configuration near the surface. The observed topological surface band dispersion is sensitive not only to the surface but also to different stackings of the underlying layers. As such, our findings point to the high tunability of both magnetic and electronic structures of the topological surface states in $(MnBi_2Te_4)_m(Bi_2Te_3)_n$. The switchable magnetic configuration and different combinations of the two building blocks would serve as two natural tuning knobs to realize various topological phases in this system.

# Methods

**Sample preparation and characterization.** Single crystals of $MnBi_2Te_4$, $MnBi_4Te_7$ and $MnBi_6Te_{10}$ were grown via a solid-state reaction method as described in an earlier study [5,40]. X-ray diffraction measurements were carried out on each sample to identify the correct phase.

**ARPES measurements.** Angle-resolved photoemission measurements were carried out on a laser-based ARPES system with a photon energy of 6.994eV. A small beam spot (~20 μm) was achieved and used to resolve different terminations in the real space of the sample surface. The samples were cleaved at 30 K in ultrahigh vacuum. A total energy resolution of ~3 meV was used for the measurements. The base pressure was better than $3 \times 10^{-11}$ mbar.

**First-principles calculations.** Our first-principles calculations based on density functional theory were performed by the Vienna *ab initio* simulation package (VASP). The exchange-correlation potential was treated within the generalized gradient approximation + Hubbard U (GGA+U) framework [41] by assuming U = 3.0 eV for Mn 3d orbitals. The experimental crystal parameters were carried out for all calculations with a cutoff energy of 350 eV for the plane wave expansion and a Gamma-centered 11 × 11 × 3 k-meshes. Spin-orbit coupling (SOC) was considered consistently in the calculations. Real-space tight-binding Hamiltonians were obtained by constructing Wannier functions for the p orbitals of Bi and Te atoms without the maximizing localization procedure using the Wannier90 package [42]. The topological properties of these tight-binding Hamiltonians were studied by the Wanniertools [43].

## Acknowledgements

We thank Y. Xu and B. Lei for useful discussions. The work at university of science and technology of China (USTC) was supported by the USTC start-up fund. The work at ShanghaiTech University was supported by the National Natural Science Foundation of China. The work at Huazhong University of Science and Technology was supported by the Ministry of Science and Technology of China (No. 2018YFA0307000) and the National Natural Science Foundation of China (No. 11874022).

## Author contributions

Y.H., L.X., M.S. and A.L. contributed equally to this work. X.-H.C. and J.-F.H. designed the research. Y.H., L.X. and S.P. carried out the ARPES measurements with support from C.F.Z., Z.K.L. and Y.L.C.. Y.H. analyzed the data with S.P. and L.X.. A.L. performed theoretical calculations with the guidance from G.X.. M.S. grew the samples with support from Z.Y.W., J.J.Y., T.W. and X.-H.C.. C.F.Z. and Z.K.L. developed and maintained the laser-ARPES facilities with support from Y.L.C.. J.-F.H. wrote the paper with inputs from all authors. J.-F.H. and X.-H.C. are responsible for project direction, planning and infrastructure.

## Additional information

**Competing interests:** The authors declare no competing financial interests.

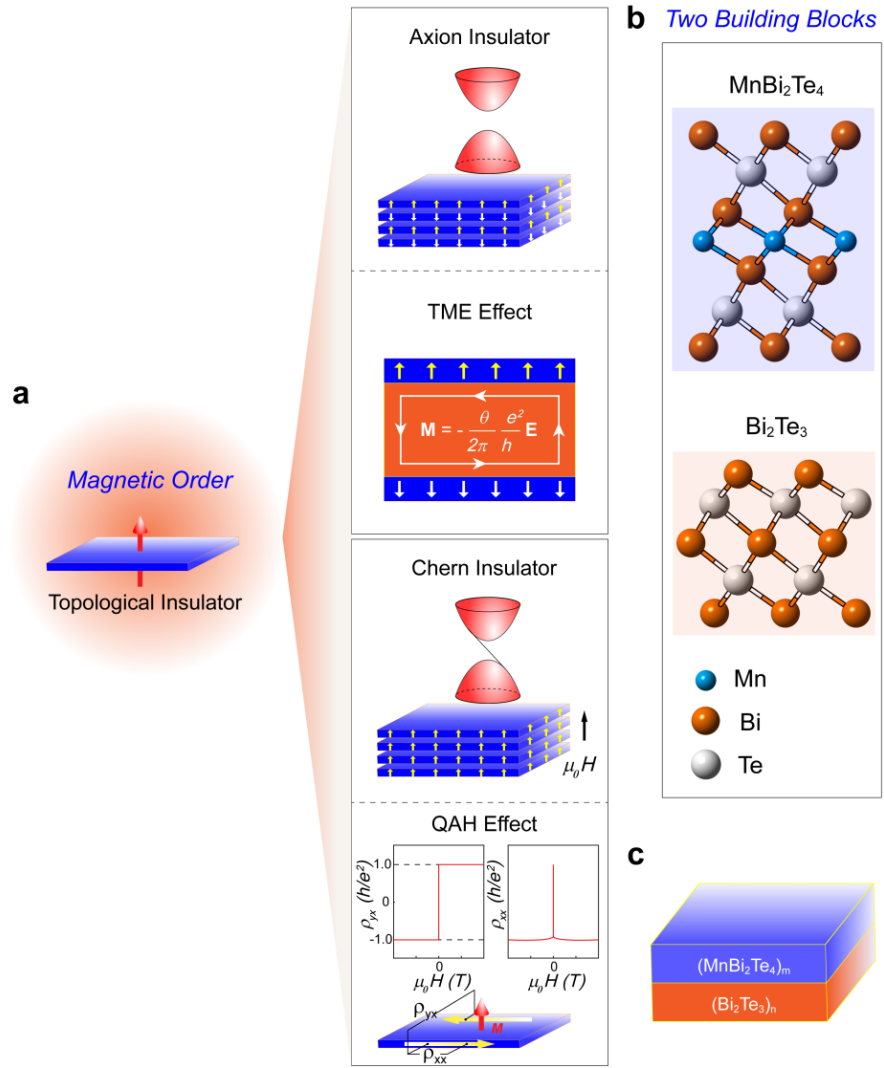

**Figure 1 | Schematics of the magnetic configurations and crystal building blocks in $(MnBi_2Te_4)_m(Bi_2Te_3)_n$ heterostructures. a**, Schematic antiferromagnetic order of the axion insulator state [26] and ferromagnetic order of the Chern insulator state [25]. **b**, Two crystal building blocks of the $(MnBi_2Te_4)_m(Bi_2Te_3)_n$ heterostructures. **c**, Different stackings of the two building blocks give rise to different materials of the $(MnBi_2Te_4)_m(Bi_2Te_3)_n$ heterostructures.

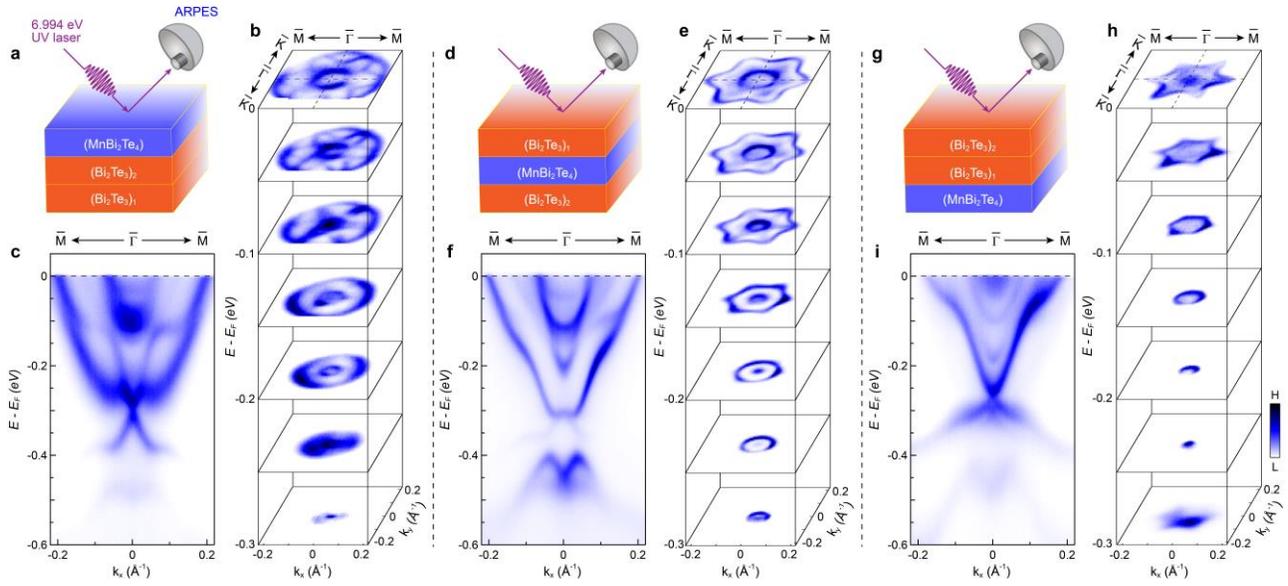

**Figure 2 | Electronic structure on different surface terminations of MnBi$_6$Te$_{10}$. a-c,** Schematic of the MnBi$_2$Te$_4$ termination (**a**), the corresponding constant energy maps (**b**) and band dispersion along the M-Γ-M high symmetry cut (**c**) measured at 6K. **d-f,** The same as **a-c,** but for type-1 Bi$_2$Te$_3$ termination as illustrated in **d. g-i,** The same as **a-c,** but for type-2 Bi$_2$Te$_3$ termination as illustrated in **g.**

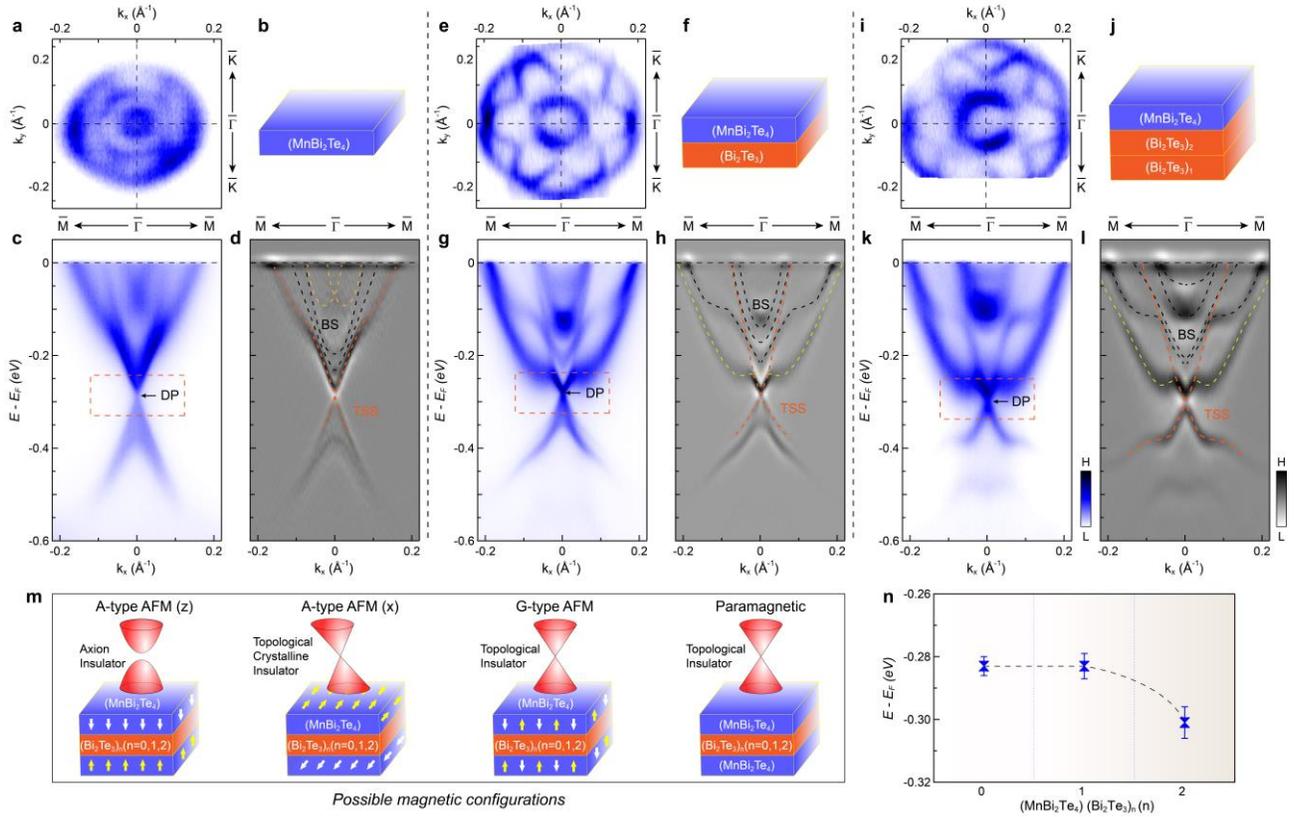

**Figure 3 | Electronic structure on the MnBi₂Te₄ terminated surfaces of MnBi₂Te₄, MnBi₄Te₇ and MnBi₆Te₁₀. a-d**, Fermi surface (**a**), schematic of the crystal building block (**b**), band dispersion (**c**) and its second derivative image along the M-Γ-M high symmetry direction (**d**) for the MnBi₂Te₄ system measured at 10K. **e-h** The same as **a-d**, but for the MnBi₄Te₇ system measured at 7K. **i-l**, The same as **a-d**, but for the MnBi₆Te₁₀ system measured at 6K. The Dirac point (DP) is emphasized by the dashed box and marked by the black arrow in **c**, **g** and **k**. The topological surface state (TSS) and bulk state (BS) are guided by the orange and black dashed lines, respectively (**d**, **h** and **l**). **m**, Schematics of different magnetic configurations and the corresponding electronic structures. **n**, Energy position of the Dirac point in different systems. The error bars represent the energy uncertainties in the determination of the Dirac point.

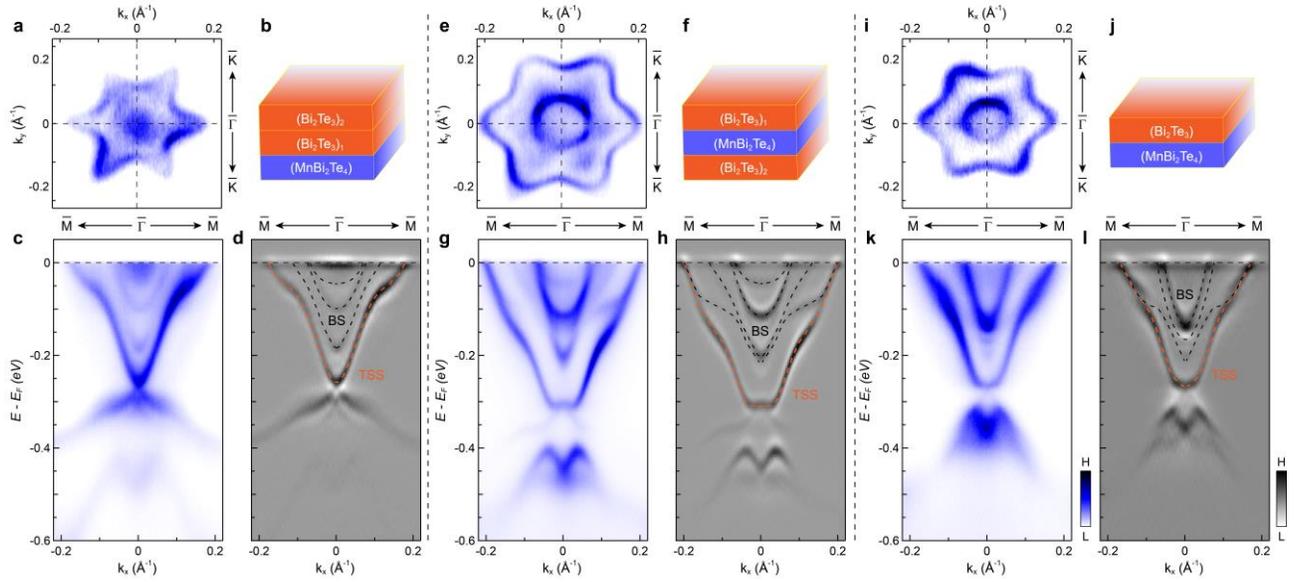

**Figure 4 | Electronic structure on the Bi₂Te₃ terminated surfaces of MnBi₄Te₇ and MnBi₆Te₁₀. a-d**, Fermi surface (**a**), band dispersion (**c**) and its second derivative image along the M-Γ-M high symmetry direction (**d**) for type-2 Bi₂Te₃ termination of MnBi₆Te₁₀, as illustrated by the schematic in **b**. **e-h**, The same as **a-d**, but for type-1 Bi₂Te₃ termination of MnBi₆Te₁₀, as illustrated by the schematic in **f**. **i-l**, The same as **a-d**, but for the Bi₂Te₃ termination of MnBi₄Te₇, as illustrated by the schematic in **j**. The measurements on MnBi₄Te₇ (MnBi₆Te₁₀) are performed at 7K (6K). The topological surface state (TSS) and bulk state (BS) are guided by the orange and black dashed lines, respectively (**d**, **h**, **l**).